# Scenarios of Destruction for Large Network and Increasing Reliability

*P. A. Golovinski* [1,2]

[1] *Moscow Institute of Physics and Technology (State University), Moscow, Russia*
[2] *Voronezh State University of Architecture and Civil Engineering, Voronezh, Russia*
*E-mail: golovinski@bk.ru*

Damage scenarios for large networks are considered. The cascade scenario is described by means of powers of adjacency matrix. More difficult probabilistic variants of the large network damage are modeling by Markov's chains. For reliability augmentation of networks we add a set of random intermediate agents with big dimensionality. It provides high reliability of all system even with low reliability of single components. Probabilistic estimation of reliability for reinforced network is made.

## 1. Introduction

Large networks usually contain many uniform elements, and that allows to apply methods of probability theory to their analysis. Damage of networks can happen under the influence of various reasons and develop according to different scenarios. Thus an important problem is allocation of often meeting scenarios. We will discuss some types of network damage. Important case is a cascade damage of network. Such damage happens when the damage of one element inevitably leads to the damage of neighboring connected element. In this case it is interesting to study dynamics of network damage of various structures. For full-connected networks, the final damage occurs after one step. This type of damage is considered in detail [1], and analytical solution of problem through adjacent matrixes of the network graph is obtained. The developed algorithm allows easily modeling a cascade destruction of connection for networks of any complexity.

## 2. Cascade destruction

We now consider damage in the form of epidemic: damage of any element leads at the following step the damage of the elements next to it. Let us take, initially, one destroyed vertex of a network and at each subsequent step it will be a damage of vertexes that are located next to the vertex destroyed earlier. If initial point is not unique, it is possible to add fictitious vertexes connected to vertexes at an initial state, and then a state of damage of the first network will arise after the first step.

Further dynamics of damaging may be visualized, if the network is small. For large systems such problem becomes unsolvable for the person, and for its solution we need to develop the adequate analytical instrument. As a result, it is desirable to define damage time for total network (number of steps until final damage), time of full loss connectivity of a network, and to plan effective tools for network protection. In the considered scenario damage evolves along adjacent vertexes of the graph. Each step of damage means an additional step on a graph in the path with length $t$. It allows one to use a known theorem to describe a network damage.

According to the theorem, the element of a matrix $A(t) = C^t$ ($C$ is adjacency matrix) is equal to number of paths of length $t$ from vertex $V_i$ to vertex $V_j$. Zero elements $A_{ij}$ shows that the vertex $V_j$ can't be reached by $t$ steps from vertex $V_i$. Final damage of a network after time $t$, when the process is started from the vertex $V_i$, means that in the corresponding row of the matrix all elements became nonzero. Let's remind that for nondirectional graph, the initial matrix $C$ is real and symmetric, both all its powers will be real and symmetric matrixes too. Thus, the matrix $A(t)$ contains all information about dynamics of damage of a network. Choosing a row, which first becomes entirely nonzero, we can define the minimum time $t_{min}$ of a network damage and initial vertex $V_i$, which corresponds to such option. Having defined the last line, which has become nonzero, we thereby will define the maximum time of damage $t_{max}$ of a network, and we will find the corresponding starting vertex.

For nondirectional graphs, owing to symmetry of a matrix $C$, it can be represented as $C = BLB^{-1}$, where $L$ is the diagonal matrix composed of eigenvalues:

$$L = \begin{pmatrix} \lambda_1 & 0 & \ldots & 0 \\ 0 & \lambda_2 & & \\ \ldots & & \ldots & \\ 0 & \ldots & & \lambda_n \end{pmatrix}. \qquad (1)$$

Then

$$A(t) = C^t = \left(BLB^{-1}\right)^t = BL^t B^{-1}, \qquad (2)$$

and

$$L^t = \begin{pmatrix} \lambda_1{}^t & 0 & \ldots & 0 \\ 0 & \lambda_2{}^t & & \\ \ldots & & \ldots & \\ 0 & \ldots & & \lambda_n{}^t \end{pmatrix}. \qquad (3)$$

For a long time, the value $(\lambda_{max})^t$ dominates.

Now we construct a matrix

$$T(t) = I + C + C^2 + \ldots + C^t \qquad (4)$$

with nonzero elements, fixing all vertexes of a network, destroyed during $t$ steps. Thus, the matrix $T(t)$ describes all the damages, that were happened in a network to the time $t$. For a plane network, it is possible to construct the isomorphic dual graph, having replaced vertexes with arches, arches by vertexes, and to consider a problem of consecutive damage of connections (arches) by the method formulated above.

Various strategies can be developed for protection of networks, depending on the purpose. If such purpose is the maximum delay of damage, it can be achieved by increasing the length of path, i.e. time $t$. From this point of view, the structure of a protection matrix means a deletion of the most quickly filled rows, i.e. a protection of the element damage will enlarge the possible path, and increase a time of final damage.

For clarifying possible options of network protection it is useful to consider some special limiting cases. It is almost impossible to protect a full-connected network with vertexes achievable by one step in cascade damage. In the radial and ring graphs the most expedient is a protecting of the central vertex with increasing a time of network damage up to time $N$ typical for vertexes on a ring. It must be kept in mind that, if protection takes vertexes out of network functioning, characteristic time of consecutive access between elements of a network increases from 2 to $N/2$. If the network is a tree, the paths of damage don't come together and final damage of a network goes long enough from any initial point of damage.

For ensuring stability of a network it is necessary to satisfy some contradictive conditions. On the one hand, the way between any two vertexes has to be the shortest. It provides fast movement of resources along a network. On the other hand, this type of connections leads to possibility of fast damage of a network. The contradiction can be eliminated if to mean two networks: one network used for normal functioning, and the second network, been reserved for functioning in a mode of protection, and being a subgraph of the first. Then the initial network under a threat of damage might be transformed to a tree. Because it may be many generating trees, it makes sense to choose the steadiest from them. The tree is characteristic, that its damage is a branching process, and the total time of damage means the longest branch, starting from initial point. Each tree has a center, i.e. vertex which is most far from the ends. For its conservation the longest way must be taken from this point $V_i$ to all possible trailing vertexes $V_j$. Then we take $V_i$, for which this value has minimum for of all vertexes:

$$t^k = \min_i \max_j \left( t_{ij}^k \right). \tag{5}$$

The index $k$ is a number of different generating columns. Further, it is necessary to choose from all graphs with final maximum value

$$\widetilde{t} = \max_k \min_i \max_j \left( t_{ij}^k \right). \tag{6}$$

The center of the final tree needs to be protected as much as possible.

### 3. Probabilistic approach

One more important class of network damage is the scenario in which damage is transferred from one vertex to the next with some probability $p_{ij}$. If consecutive damages are mutually independent, the dynamics of a process may be represented by Markov's chain [2]. The corresponding ergodic classes show domain of the guaranteed final damage. In continuous description, it can be represented in the form of the differential equation

$$\frac{d\pi(t)}{dt} = \pi(t)A,$$ (7)

where $\pi(t)$ is a vector of a network state, describing the probability of vertex damage at a time $t$, $A = \{a_{ij}\}$ is a matrix of a transition rate. If the initial state of a network is described by a vector of state $\pi(0)$, the solution to the Eq. (7) has the form

$$\pi(t) = \pi(0)e^{At}.$$ (8)

The maximum eigenvalue of a matrix $A$ defines an average number $\lambda$ of connections for one vertex. It allows one to estimate the average rate of damage, and the decreasing number of connections at cascade destruction is proportional to $\exp(-\lambda t)$ at the beginning of damage. In a random graph, the probability for all connections are identical. Conservation of connectivity for a random graph is dictated by Erdos-Renyi [3] theorem: if probability of vertex connection of the full-connected graph $p = c\ln n / n$, and $c > 3$, the network remains full-connected with probability

$$P \geq 1 - 1/n, \ n \to \infty.$$ (9)

Conservation of connections and restoration opportunity are of great importance for networks of different types, including the neural network systems of associative memory, basing on dynamic Hopfild's networks [4]. As shown in [5], these networks are quite stable with respect to partial damage, keeping ability for restoration of associative connection at a survival of only 30% of communications. The disadvantage of such networks is their full-connectivity, which leads to instability of a network, mentioned above, when we looked to the problem of cascade damage.

Further increasing of reliability for associative networks, as well as for some information systems, can be reached by adding a connection with an auxiliary set of elements of higher dimension $N$ [6]. We discuss a system with two input sets $X$ and $Y$, which dimensions are $r$ and $k$, and the projective frontal set (layer) $A$, containing $N \gg r, k$ elements. Elements of vectors are connected to elements of a frontal set in a random way with probability $p$. The dimensions are fixed as $r$ and $k$, and the dimension $N$ is large and can be changed. We consider, how formation of connection between sets $X$ and $Y$ through a frontal set $A$ depends on the probability $p$ of connection formation and dimension $N$.

For the next step, we choose any vertex $V_i$ from a frontal layer. Probability that it isn't connected to one of the vertexes $X$ is equal $(1-p)$, and for all vertexes of this entrance layer it is equal $(1-p)^r$. The probability that vertex $V_i$ is connected at least to one point from $X$ is equal $1-(1-p)^r$. The same reasoning concerns to connection with vertex from a set $Y$. The probability that vertex $V_i$ is connected at least to one vertex from $Y$ is equal $1-(1-p)^k$. The probability that vertex $V_i$ is connected at least to one vertex from $X$, and to one from $Y$ is equal

$$p_c = \left(1-(1-p)^r\right)\left(1-(1-p)^k\right). \qquad (10)$$

We assumed that probabilities of connection of vertexes are independent from each other. In this case the probability of connection of an element from $X$ with an element from $Y$ can be found as a product of probabilities. If we assume, at the same time, $p \to 0$ and $N \to \infty$ at fixed $r$ and $k$ $(r < k)$, then the probability of connection $X$ and $Y$ through $A$ is of form

$$\left(1-(1-p)^r\right)\left(1-(1-p)^k\right) \approx (rp)(kp) + O(p^3). \qquad (11)$$

It also doesn't depend on a choice of $V_i$. For the next, we take $P = (Nr)^{-1/2}$. The average number of vertexes, activated in a frontal set, will be

$$N_a = Np_c = Nrk\left(\frac{1}{Nr} - O\left(N^{-3/2}\right)\right) = k - \left(N^{-3/2}\right). \qquad (12)$$

The variation of a connection number is equal

$$N_a = Np_c(1-p_c) = k - O\left(N^{-1/2}\right). \qquad (13)$$

Thereby, the random graph with $P = (Nr)^{-1/2}$ contains average number of connections $k$ that completely provides a connection of inputs with outputs through a frontal layer. The randomly organized connections form intersection in domain $A$: $F(X) \cap F(Y) \subset A$ that is equivalent to logical operation $X \wedge Y$.

On the basis of consecutive system of similar connections, it is possible to build up hierarchical logical system. Reliability of a system is provided by repeated generating of the random graphs after damage of $k$ vertexes from $N$ in $A$, so then again will arise $k$ connections instead of the lost.

## 4. Conclusions

One of the effective ways of connection conservation in a complex network is exploitation of large number additional intermediate agents with low reliability that is compensated by their number. The sparse system with random connections through an auxiliary set, having very large number of elements, can be used repeatedly with insignificant probability of overlapping.

The work was supported by RFFR  (project No. 11-07-00155-a).